\documentclass[twocolumn,english,prl]{revtex4-1}
\usepackage[latin9]{inputenc}
\usepackage{amsmath}
\usepackage{amssymb}
\usepackage{graphicx}
\usepackage{babel}
\begin{document}
\global\long\def\ket#1{\left|#1\right\rangle }

\global\long\def\bra#1{\left\langle #1\right|}

\global\long\def\braket#1#2{\left\langle #1\left|#2\right.\right\rangle }

\global\long\def\ketbra#1#2{\left|#1\right\rangle \left\langle #2\right|}

\global\long\def\braOket#1#2#3{\left\langle #1\left|#2\right|#3\right\rangle }

\global\long\def\mc#1{\mathcal{#1}}

\global\long\def\nrm#1{\left\Vert #1\right\Vert }

\title{Markovian heat sources with the smallest heat capacity}

\author{Raam Uzdin}

\affiliation{Technion - Israel Institute of Technology, Haifa 3200008, Israel}

\author{Simone Gasparinetti}

\affiliation{Department of Physics, ETH Zurich, CH-8093 Zurich, Switzerland}

\author{Roee Ozeri}

\affiliation{Department of Physics of Complex Systems, Weizmann Institute of Science,
Rehovot 7610001 Israel}

\author{Ronnie Kosloff}

\affiliation{Hebrew University of Jerusalem, Jerusalem 9190401, Israel}

\begin{abstract}
Thermal Markovian dynamics is typically obtained by coupling a system
to a sufficiently hot bath with a large heat capacity. Here we present
a scheme for inducing Markovian dynamics using an arbitrarily small
and cold heat bath. The scheme is based on injecting phase noise to
the small bath. Several unique signatures of small bath are studied.
We discuss realizations in ion traps and superconducting qubits and
show that it is possible to create an ideal setting where the system
dynamics is indifferent to the internal bath dynamics.
\end{abstract}
\maketitle
Thermodynamics of small system has been intensively studied in recent
years. Stochastic thermodynamics explores the relations between different
trajectories in the system phase space \cite{Seifert2012StochasticReview,harris2007fluctuationReview}.
Quantum thermodynamics \cite{Goold2015review,SaiJanetReview,millen2015review}
deals with the effects of non-classical dynamics and non-classical
features such as coherence and entanglement on thermodynamics. Apart
from the practical importance of understanding and experimenting with
thermodynamics at the smallest scales, the study of quantum thermodynamics
has also provided exciting theoretical developments. As an example,
it has been shown that there are additional second law-like constraints
on small systems interacting with a thermal bath \cite{GourRTreview,LostaglioRudolphCohConstraint,RUgenCluasius2ndLaws}.
In addition there are thermodynamic effects in heat machines that
can be observed only in systems that are sufficiently ``quantum''
\cite{MitchisonHuber2015CoherenceAssitedCooling,EquivPRX,RUcollective,Wlofgang2016nonThermalBaths}.
Furthermore, some quantum heat machine setups can exceed classical/stochastic
bounds \cite{EquivPRX,RUcollective}.

In a self-contained nanoscale setup (e.g. ion traps) that includes
the baths, the heat capacities of the baths will be nanoscopic as
well. How small can an object be and still perform as a thermal bath?
What are the features of a large bath that a microscopic heat bath
can mimic? The most pronounced feature of an large ideal bath is its
lack of memory. Given the state of the system and the bath temperature,
the change in the state of the system is fully determined. Moreover,
the final state of the system will be a thermal state with the original
bath temperature. To achieve this, the bath has to be very large and
sufficiently hot to ensure a short correlation time with respect to
the system bath coupling strength \cite{breuer}. In this paper we
suggest a scheme for a bath that generates Markovian dynamics for
1) arbitrary low temperature, and 2) arbitrary small heat capacity. 

\subsection*{General setup}

Our setup is composed of a system, a small bath (energy reservoir)
and an external dephasing source for the small bath. The small bath
is an ensemble of N spins (qubit, or other two-level systems). Excluding
the external dephasing source the Hamiltonian that describes our setup
is

\begin{equation}
H_{tot}=H_{0}+V=h_{0}\sum_{i}^{N+1}\sigma_{i}^{z}+\sum_{i>j}^{N+1}V_{ij},\label{eq: Hb_tot}
\end{equation}
where $\sigma_{i}^{z}$ is the Pauli z matrix of spin $i$. The first
$N$ spins constitute the small bath, and spin $N+1$ is the system.
Alternatively, the system can be larger but the bath interacts only
with two levels in the system as often the case in quantum heat machines
\cite{scovil59,EquivPRX,k102}. The spin-spin energy conserving interactions
(between the system and the bath or between the bath spins) are of
the form 
\begin{equation}
V_{ij}=\xi_{ij}(\sigma_{i}^{-}\sigma_{j}^{+}+\sigma_{i}^{+}\sigma_{j}^{-}),\label{eq: Hint}
\end{equation}
where $\sigma_{i}^{+(-)}$ is the creation (annihilation) operator
of spin $i$. The first configuration we study is ``all to all''
(ATA) coupling, where all spins are equally coupled to each other
$\xi_{ij}=\xi$. The second configuration is a linear chain with nearest
neighbor (NN) coupling $\xi_{ij}=\xi_{ij}\delta_{i,j+1}.$ These configurations
can be implemented in ion traps and superconducting circuits as discussed
in the end of the paper.

When $N$ is a small number and there is no dephasing, frequent quantum
recurrences takes place and energy oscillates back and forth between
the system and the bath. In general the system will not relax to a
steady state.

\subsection*{Dephased baths - adding an additional dephasing environment}

In our setup each \textit{bath} spin ($1$ to $N$) is subjected to
dephasing (phase damping) created by some larger environment. Strong
dephasing can be an intrinsic, possibly tunable, property of the spin
system, or it can be artificially induced by noise engineering. The
spin decoherence dynamics is described by a Lindblad equation as described
in the Appendix. In the main text it will suffice to denote the coherence
relaxation rate by $\alpha$. The dephasing can be replaced by repeated
projective energy measurements ($\sigma^{z}$) of the spins in the
bath. The typical time between subsequent measurements will determine
the effective dephasing rate $\alpha$.

The decoherence is generate by another external environment (not the
$N$-spin bath). Yet, we consider it as a free resource for the following
reasons. First, unlike thermalization, dephasing is very often easy
to engineer or to add to existing schemes. Second, dephasing does
not change the energy distribution so it cannot generate work or heat
flows. That is, the dephasing environment is energetically useless.
In our setup the energy exchange with the system comes only from the
small bath (spins $1$ to $N$). Third, the dephasing environment
can only increase the entropy of the elements it interacts with -
it cannot be used as a resource for entropy reduction. 

Our scheme is different from other schemes where the system is directly
dephased \cite{Gershon2000NatuteZenoAZ,Gershon2008thermoZeno,Gershon2011zeno}.
There, the coherence of the system degrades at a rate that is much
faster than the thermalization rate. Having roughly comparable dephasing
and thermalization rates, is highly important for observing certain
quantum thermodynamics effects (e.g. \cite{EquivPRX,MitchisonHuber2015CoherenceAssitedCooling}).
Interestingly, classical noise has recently been used to simulate
quantum Markovian dynamics \cite{Chenu2017ClassicalNoiseLidblad}.
There the classical noise acts as an infinite heat capacity bath.
Hence, \cite{Chenu2017ClassicalNoiseLidblad} is very different from
the present study. While our scheme can lead to Markovian dynamics
for the system, it is shown that some features of the bath smallness
can still be observed. These small bath features have not been explored
before to the best of our knowledge. 

Dephasing eliminates coherence between energy eigenstates of each
bath spin, but it also eliminates inter-particle coherence between
spins (off diagonal element in the multi-particle density matrix in
the energy basis). We show that this inter-particle coherence mitigation
prevents recurrences and leads to Markovian dynamics. Let $z_{s}$
denote the polarization of the system spin $z_{s}=tr[\sigma_{N+1}^{z}\rho]$.
In the Appendix we find that for the ATA and NN configurations, strong
dephasing (NN also requires weak system-bath coupling) leads to the
following equation of motion

\begin{align}
\frac{d^{2}}{dt^{2}}z_{s} & =2\xi_{SB}^{2}\frac{N+1}{N}(z_{\infty}-z_{s})-\alpha\frac{d}{dt}z_{s},\label{eq: d2z no approx}\\
z_{\infty} & =\frac{\sum_{i=1}^{N+1}z_{i}(0)}{N+1}=\frac{N}{N+1}z_{T}+\frac{1}{N+1}z_{s}(0),
\end{align}
where $z_{T}$ is the initial average polarization of the bath spins,
and it is equal to $z_{T}=-\tanh[h_{0}/(2T)]$ when the bath is prepared
in thermal state of temperature $T$. In deriving (\ref{eq: d2z no approx})
we have neglected terms of order $O(\left\langle \sigma_{j}^{+}\sigma_{k}^{z}\sigma_{l}^{-}\right\rangle +c.c.).$
These terms are proportional to coherences between particles which
is strongly suppressed by the external dephasing (see Appendix). Hence,
the reduced dynamics is described the second order equation (\ref{eq: d2z no approx}).
For the ATA case $\xi_{SB}^{2}=N\xi^{2}$ and for the NN configuration
$\xi_{SB}^{2}=\xi_{N,N+1}^{2}$. 

$z_{\infty}$ is the polarization of the system after the system and
bath fully equilibrate to a state where all the $N+1$ spins have
the same polarization. The dependence of $z_{\infty}$ on the system
initial state is a direct consequence of having a bath with finite
energy (small heat capacity).

For the system coherence, for example $x_{s}=tr[\sigma_{N+1}^{x}\rho]$,
the situation is somewhat simpler since the bath has no initial coherence
and we get
\begin{equation}
\frac{d^{2}}{dt^{2}}x_{s}=\xi_{SB}^{2}x_{s}-\alpha\frac{d}{dt}x_{s}.\label{eq: d2 x}
\end{equation}

\subsection*{General features of the reduced dynamics }

We start by looking at different regimes of operation depending on
the value of $\alpha/\xi_{SB}$. Equation (\ref{eq: d2z no approx})
reveals that the only solution consistent with strong dephasing $\alpha/\xi_{SB}\to\infty$
is the Zeno freeze-out $\frac{d}{dt}z_{s}(t)=0$. This can be useful
to decouple the system from the bath without changing $\xi_{SB}$.
The next regime of interest is Markovian dynamics. As $\alpha$ gets
smaller and more comparable to $\xi_{SB}$ the system starts to evolve
in a Markovian manner. By neglecting the second derivative in (\ref{eq: d2z no approx})
we get a Markovian equation, and its solution is

\begin{equation}
z_{s}^{Mark}=z_{\infty}+(z_{s}(0)-z_{\infty})\exp[-\frac{2\xi_{SB}^{2}}{\alpha}\frac{N+1}{N}t].\label{eq: A2A mark}
\end{equation}
In accordance with the Zeno freeze-out the thermalization rate satisfies
$\frac{\xi_{SB}^{2}}{\alpha}\frac{N+1}{N}\to0$ when $\alpha\ggg\xi_{SB}$.
By evaluating $\frac{d^{2}}{dt^{2}}z_{s}^{Mark}$ and dividing by
$\xi_{SB}^{2}$ we find that it is $O[(\frac{\xi_{SB}}{\alpha})^{2}]$
whereas the other terms are $O[(\frac{\xi_{SB}}{\alpha})^{0}]$. Thus,
Markovian dynamics is observed when the dephasing is sufficiently
larger than the system-bath coupling strength ($\alpha\gg\xi_{SB}$).
In practice, when the dephasing rate is roughly ten times larger than
the coupling coefficient, the dynamics is already highly Markovian. 

For a bath initially prepared at temperature $T$, a system in a thermal
state with temperature $T$ is a \textit{fixed point} of the setup.
The \textit{asymptotic state} is the state that a non-thermal state
will reach after a very long time with respect to the thermalization
time. In large baths the asymptotic state and the fixed point are
the same. This is different in small baths. While the thermal state
is still a fixed point of thermalization maps, the asymptotic state
is given by $z_{\infty}$ and not by the polarization of thermal state
$z_{T}$ determined by the initial temperature of the bath. From the
definition of $z_{\infty}$ one can verify that in the large bath
limit $N\gg1$, $z_{\infty}\to z_{T}$ and the asymptotic state becomes
equal to the fixed point.

The large $\alpha/\xi_{SB}$ limit of (\ref{eq: d2 x}) leads to Markovian
dynamics for the coherence $x_{s}^{Mark}=x_{s}(0)\exp[-\frac{\xi_{SB}^{2}}{\alpha}t]$.
Comparing the exponential decay rate of the coherence and polarization
we find $\gamma_{z}=2\frac{N+1}{N}\gamma_{x}$. As shown in the Appendix,
the $\frac{N+1}{N}$ enhancement of the decay rate with respect to
twice the dephasing rate, \textit{is an effect unique to small heat
capacity baths}. This enhancement is not in contradiction to the known
$\gamma_{z}\le2\gamma_{x}$ ($T_{2}\le2T_{1}$) relation, valid for
completely positive \textit{time-independent} Markovian maps \cite{gorini76,chang1993T1T2}.
The $\frac{N+1}{N}$ enhancement, is due to non-negligible changes
in the average populations in the bath (for $N=O(1)$). See Appendix
for further details. This enhanced decay rate is an experimentally
measurable signature of small dephased baths.

\subsection*{Closed form non-Markovian reduced dynamics}

Starting at $t=0$ without system-bath coherence implies $\frac{d}{dt}z(0)=0$.
Thus, at early times before the system starts to follow Markovian
dynamics the second derivative in (\ref{eq: d2z no approx}) dominates.
This second derivative is a reminiscent of the unitary dynamics that
takes place in the absence of dephasing. Unlike more complicated setups
with strong memory effects, here all the non-Markovian effects are
encapsulated in the second derivative term. A numerical example is
given below.

\subsection*{The ATA configuration}

In this configuration $\xi_{SB}^{2}=N\xi^{2}$. The $N$ factor in
$\xi_{SB}^{2}$ is expected since the system is connected to $N$
spins. In the numerical example in Fig. 1 we use a three spin bath
that interacts with a system (another spin) via ATA coupling. The
parameters are $\xi=1$ and $\alpha=6$. The dashed black curve in
Fig. 1a shows the Markovian approximation (\ref{eq: A2A mark}) with
respect to the exact dynamics (red curve). A most appealing feature
of the ATA configuration is that the reduced dynamics of the system
does not depend on the internal polarization distribution in the bath.
Only the total polarization of the bath (total energy) affects the
system (see Appendix for more information). This means that to get
a thermalization dynamics there is no need to carefully prepare the
bath in a thermal state where all the spins are uncorrelated and have
the same polarization. This feature is a major simplification both
for practical (or experimental) considerations and for theoretical
considerations. The green-dashed curves in Fig. 1a shows that a completely
different bath preparation with the same initial total polarization,
leads to the exact same system dynamics (analytically the same so
it is not visible in the graph). Moreover, even strong classical correlation
in the bath (e.g. $(1-p)\left|000\right\rangle \left\langle 000\right|+p\left|111\right\rangle \left\langle 111\right|$)
will not effect the reduced dynamics of the system. In Fig. 1b the
free evolution without external dephasing is plotted. Quantum recurrences
dominate the dynamics and equilibrium is not achieved.

Figure 1c shows that (\ref{eq: d2z no approx}) accurately describes
the short-time non-Markovian evolution. For short time evolution the
second derivative in (\ref{eq: d2z no approx}) is highly important
even if $\alpha\gg\xi_{SB}$. 

For a large bath without dephasing weak system-bath coupling is crucial
for observing Markovian dynamics. The coupling has to be smaller than
the bath correlation time, which is proportional to the bath temperature
\cite{breuer}. As a result the Lindblad description fails for very
cold baths. In contrast, \textit{In our setup there is no such limitation}.
The system bath coupling has to be small compared to the dephasing
rate. Under sufficiently strong dephasing the system will follow Markovian
dynamics \textit{regardless of how cold} is the initial temperature
of the bath.
\begin{figure}
\includegraphics[width=8.6cm]{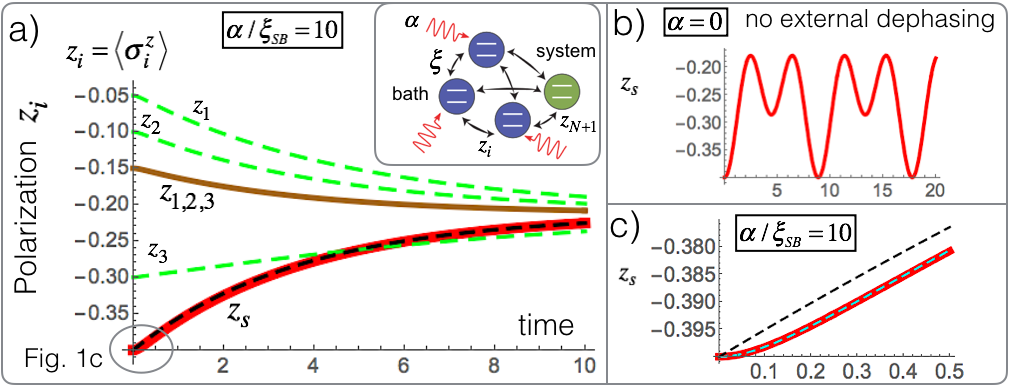}

\caption{(a) The red curve shows the (exact) polarization of the system $z_{s}$
as a function of time, while the brown solid line stands for the polarizations
of the bath spins when starting from a uniform polarization state
($z_{1}=z_{2}=z_{3})$. The dashed-black line shows the Markovian
approximation (\ref{eq: A2A mark}) with respect to the exact dynamics
when the dephasing is stronger than the system-bath coupling $\alpha/\xi_{SB}=10$.
In the 'all to all' coupling completely different initial bath preparations
with the same initial total polarization (green dashed lines) lead
to the exact same system dynamics (red curve). Thus, in this ideal-like
bath the only important parameter of the bath is the total energy.
(b) When the external dephasing is off ($\alpha=0$) equilibrium is
not reached due to quantum recurrences. (c) For very short evolution
time the Markovian approximation (dashed-black (\ref{eq: A2A mark}))
fails but our non-Markovian correction (dashed-cyan (\ref{eq: d2z no approx}))
accurately describes the evolution. }

\end{figure}

\subsection*{Weak vs. strong coupling in dephased baths }

When the coupling of the system to the bath $\xi_{SB}$ is much weaker
than the coupling between different spins in the bath $\xi_{ij}^{B}$
($i,j<N+1)$ the dynamics greatly simplifies. In the presence of dephasing
and $\xi_{ij}^{B}\gg\xi_{SB}$, the bath spins equilibrate among themselves
before the system changes significantly. Thus, (\ref{eq: d2z no approx})
is valid also for configuration such as NN with weak system-bath coupling
(see Appendix). In the Appendix we write an equation similar to (\ref{eq: d2z no approx})
for the NN configuration when $\xi_{ij}^{B}\gg\xi_{SB}$ does not
hold. 

\subsection*{System-bath correlation in small dephased baths }

Next we study the creation of system-bath correlation during Markovian
and non-Markovian dynamics of the system. When the system is weakly
coupled to a \textit{large} bath, the correlation between the system
and the bath can be made negligible and be ignored. This is not the
case for small baths. In the following we discuss both the NN and
the ATA configurations. As an illustrative example in the NN setup
we use three spins for the bath with $\alpha=6$, $\xi_{B}=1$ and
$\xi_{SB}=1/20$. For the ATA setup we also use three spins for the
bath and $\alpha=6,\xi=1/(20\sqrt{3})$. As a result, the Markovian
decay time for both configurations is the same, and is equal to $\tau_{MARK}=\alpha/(2\xi_{SB}^{2})=1200$.
The top curves in Fig. 2a show that the system polarization in the
NN configuration (red curve), and the polarization in the ATA configuration
(dashed blue) have practically the same evolution. The tiny difference
arises from the fact that the bath spins (lower curves show $z_{1},z_{2},z_{3}$)
do not yet perfectly equlibriate among themselves. Correction to Markovian
dynamics are observed only on a scale of $1/\alpha=1/6$. These correlations
are classical since off-diagonal elements are negligible in the presence
of strong dephasing. We study the correlation by looking at the standard
statistical correlation $\text{corr}(i,j)=\text{cov}(z_{i},z_{j})/[\text{var}(z_{i})\text{var}(z_{i})]$.
The dashed-blue curve in Fig. 2b shows the correlation between any
of the bath spins and the system in the ATA configuration. A more
interesting dynamics takes place in the NN setup. Remarkably, even
though the bath spins in the NN case have almost exactly the same
polarizations (Fig. 2a), their correlation with the system differ
significantly at $1/\alpha\le t\ll\tau_{MARK}$ as shown by the red
curves in Fig. 2b (see the inset for a magnification). This highly
interesting correlation equilibration at a rate much slower than the
polarization equilibration warrants further study. 

Another interesting feature in the correlation evolution is the asymptotic
value obtained for $t\gg\tau_{MARK}$. While the transient correlation
evolution depends on the coupling strength and on the coupling configuration,
the asymptotic value $\mathit{\text{corr}{}_{\infty}}$ depends only
on the initial conditions. This can be understood from the fact that
$\sum_{i\neq j}^{N+1}\left\langle \sigma_{i}^{z}\sigma_{j}^{z}\right\rangle $
is a conserved quantity in for choice of $\xi_{ij}$. Using this conserved
quantity together with energy conservation, and the fact that the
final equilibrium state is completely uniform, we obtain the asymptotic
correlation $\mathit{\text{corr}{}_{\infty}}=[\frac{1}{(N-1)N}\sum_{i\neq j}^{N+1}\left\langle \sigma_{i}^{z}\sigma_{j}^{z}\right\rangle _{t=0}-z_{\infty}^{2}]/(1/4-z_{\infty}^{2}).$
When the spins are initially uncorrelated and the bath starts in a
uniform polarization this expression simplifies to

\begin{equation}
\mathit{\text{corr}{}_{\infty}}=-\frac{(z_{T}-z_{s}(0))^{2}}{(M+1)^{2}}/(1/4-z_{\infty}^{2})\label{eq: corr inf}
\end{equation}
Note that regardless of the configuration, at $t\to\infty$ the system
is equally correlated to \textit{all} bath spins. $\text{corr}{}_{\infty}$
is negative for any $M$, $z_{T}$, and $z_{s}(0)$. Moreover, since
$z_{\infty}\to z_{T}$ for large $N$, we conclude from (\ref{eq: corr inf})
that the asymptotic correlation scales like $1/(M+1)^{2}$ when $z_{T}$
and $z_{s}(0)$ are kept fixed. Hence, $\mathit{\text{corr}{}_{\infty}}\to0$
for large bath. We conclude that the unavoidable asymptotic system-bath
correlation is another unique feature that appears in small dephased
bath, even when the dynamics is fully Markovian. In small \textit{non-dephased}
baths the correlation will also be strong but the correlation will
oscillate without reaching a steady state.

\begin{figure}
\includegraphics[width=8.6cm]{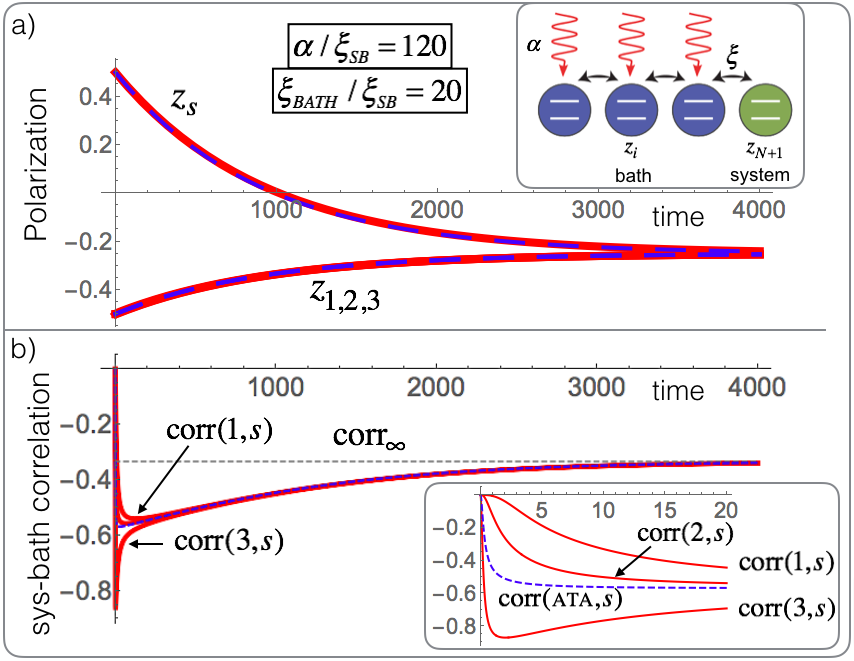}

\caption{In our scheme is strong system-bath correlation can form even when
the system follows Markovian dynamics. (a) In the weak coupling limit
in the NN configuration (see inset), the polarization dynamics (red
curves) is the same as that of the ATA configuration (blue dashed
curves) when the ATA configuration is set to have the same decay time
$\alpha/(2\xi_{SB})^{2}$. Yet, the system-bath correlation (b) in
the NN setup (red) is non-uniform even when the bath polarization
is highly uniform (see inset). The the ATA correlation is shown in
the dashed-blue curve. While the peak correlation depends on the setup
parameters, the unavoidable asymptotic correlation $\text{corr}{}_{\infty}$
(\ref{eq: corr inf}) is fixed by the initial conditions and cannot
be mitigated by weak system-bath coupling. }

\end{figure}

\subsection*{Experimental realization in ion traps }

Quantum simulation of spin Hamiltonians has been demonstrated using
trapped atomic ions \cite{Schaetz2008,Monroe2010,Roos2014}. An effective
spin 1/2 system (pseudo-spin) is realized using two internal levels
of the ion which can be separated by radio-frequency, microwave or
optical transitions. By using spin dependent forces that couple internal
and translational degrees of freedom, it is possible to generate various
interactions between the ions. In the Appendix we discuss NN and ATA
realization in ion traps, and find that it should be possible to implement
a dephased bath scheme with the current experimental capabilities.

\subsection*{Experimental realization superconducting circuits}

dephased baths can be also readily implemented in superconducting
circuits. By coupling transmon-type superconducting qubits \cite{Koch2007}
to a common far detuned resonator the NN and the ATA configurations
can be implemented. In the Appendix we provide estimates and mention
previously implemented building blocks. The superconducting architecture
also allows to experimentally explore possible extensions of this
work, including studies of system-bath correlations, continuously
driven open system \cite{Gasparinetti2013}, etc.

\subsection*{Conclusion}

We have introduced a paradigm for the implementation of Markovian
heat sources with the smallest possible heat capacity. Despite the
Markovian dynamics we have identified features in our setup that are
unique to small heat sources: 1) An enhanced decay rate with respect
to the decoherence rate 2) A significant, and unavoidable system-bath
correlation that cannot be eliminated by reducing the system-bath
coupling. In addition, we find a correlation equilibration time that
is much larger than the bath polarization equilibration time. Finally,
our theoretical approach also provides an accurate reduced description
for the non-Markovian dynamics at short evolution times. 

We have studied realizations in ion traps and in superconducting qubits
where the ``all to all'' coupling can be implemented. In our dephased
bath setup the ATA coupling has the remarkable property that the system
is sensitive only to the \textit{total} energy in the bath. This makes
the ATA bath an ideal bath: the internal dynamics inside the bath
does not affect the reduced dynamics of the system. This greatly simplifies
the preparation stage of the bath: any preparation with the same energy
will do.

Potentially, these baths can serve as a practical element in experiments.
Furthermore, this setup motivates new questions about open quantum
systems. For example, the study of accumulated system-bath correlation
is a complicated topic that is greatly simplified in the dephased
bath setup proposed here. In addition, it is intriguing to study dephasing
in quantum system with larger Hilbert space (qudits) and to include
system-bath particle exchange.
\begin{acknowledgments}
This work was supported by the Israeli science foundation. The work
of RO is supported by grants from the ISF, I-Core, ERC, Minerva and
the Crown photonics center. SG and RU are grateful to J. Heinsoo and
Y. Salathé for useful discussions.
\end{acknowledgments}

\section*{Appendix}

\subsection*{Derivation of the equation of motion}

The system-bath setup dynamics is modeled by a Lindblad equation
\begin{equation}
d_{t}\rho=-i[H,\rho]+\sum2A_{k}\rho A_{k}^{\dagger}-A_{k}^{\dagger}A_{k}\rho-\rho A_{k}^{\dagger}A_{k},\label{eq: Lindblad}
\end{equation}
where $A_{k}$ describes the external dephasing environment, and $H$
is given by Eq. (1) in the main text. By moving to the interaction
picture $H_{0}$ is eliminated and $V$ remains unaffected Since we
$[H_{0},V]=0$. To obtain a dephasing rate $\alpha$ for spin $i$
we set $A_{k}=\sqrt{\alpha}\sigma_{i}^{z}$ . Note that for two-spins
$A_{1}$ (or $A_{2}$) will also lead to dephasing at a rate $\alpha$
of coherences that involve both spins for example element $\rho_{2,3}$
in the joint density matrix.

We start the derivation of the equation of motion by writing the equation
for the time derivative of polarization of the system spin $z_{s}=\left\langle \sigma_{s}^{z}\right\rangle $

\begin{align}
d_{t}z_{s} & =\text{tr}\{-i\sigma_{s}^{z}[V,\rho]\nonumber \\
 & +\sigma_{s}^{z}\sum2A_{k}\rho A_{k}^{\dagger}-A_{k}^{\dagger}A_{k}\rho-\rho A_{k}^{\dagger}A_{k}\}.
\end{align}
Since $[A_{k},\sigma^{z}]=0$
\begin{align*}
d_{t}z_{s} & =\text{tr}[-i\sigma_{s}^{z}(V\rho-\rho V)]=-i\text{tr}([\sigma_{s}^{z},V]\rho).
\end{align*}
In contrast to other methods for obtaining reduced dynamics that are
based on integration (see the microscopic derivation \cite{breuer}),
we start with differentiation

\begin{align}
d_{t}^{2}z_{s} & =\text{tr}\{-i\sigma_{s}^{z}[H,d_{t}\rho]\nonumber \\
 & +\sigma_{s}^{z}\sum2A_{k}d_{t}\rho A_{k}^{\dagger}-A_{k}^{\dagger}A_{k}d_{t}\rho-d_{t}\rho A_{k}^{\dagger}A_{k}\}\\
 & =\text{tr}\{-\sigma_{s}^{z}[V,[V,\rho]]\}\nonumber \\
 & -\alpha\text{tr}\{i\sigma_{s}^{z}[V,(\sum_{k}2\sigma_{k}^{z}\rho\sigma_{k}^{z}-\frac{1}{2}\rho)\}.\label{eq: d2 first stage}
\end{align}
Let us first study the first term (quadratic in $V$)
\begin{align}
\text{tr}\{\sigma_{s}^{z}[V,[V,\rho]]\}= & \text{tr}\{\rho[[\sigma_{s}^{z},V],V]\}\nonumber \\
= & \text{tr}\{\rho[[\sigma_{s}^{z},\sum_{k}^{N}V_{s,k}],\sum_{j=1}^{N+1}V_{k,j}]\}\nonumber \\
= & \sum_{k}^{N}\text{tr}\{\rho[[\sigma_{s}^{z},V_{s,k}],V_{s,k}]\}\nonumber \\
+ & tr\{\rho[[\sigma_{s}^{z},\sum_{k}^{N}V_{s,k}],\sum_{j=1}^{N}V_{k,j}]\}\nonumber \\
= & \sum_{k=1}^{N}2(z_{s}-z_{k})\xi_{sk}^{2}\nonumber \\
+ & 2\sum_{k,j=1}^{N}\xi_{sk}\xi_{sj}(\left\langle \sigma_{k}^{-}\sigma_{j}^{z}\sigma_{s}^{+}\right\rangle +c.c.)\nonumber \\
+ & 2\sum_{k,j=1}^{N}\xi_{sk}\xi_{sj}(\left\langle \sigma_{k}^{-}\sigma_{s}^{z}\sigma_{j}^{+}\right\rangle +c.c.).\label{eq: unit part}
\end{align}
Next we study the second term in (\ref{eq: d2 first stage}) (Linear
in $\alpha$)
\begin{align}
\alpha\text{tr}\{i\sigma_{s}^{z}[V,(\sum_{k}2\sigma_{k}^{z}\rho\sigma_{k}^{z}-\frac{1}{2}\rho)\} & =\alpha i\text{tr}\sum_{k}\rho\sigma_{k}^{z}[\sigma_{s}^{z},V]\sigma_{k}^{z}\nonumber \\
 & =-i\alpha\text{tr}\sum_{k}\rho[\sigma_{s}^{z},V]\nonumber \\
 & =\alpha d_{t}\left\langle z\right\rangle ,
\end{align}
where we have used the relation $\sigma^{z}\sigma^{\pm}\sigma^{z}=-\frac{1}{4}\sigma^{\pm}$
. Finally we get
\begin{equation}
\frac{d^{2}}{dt^{2}}z_{s}=-2\sum\xi_{sk}^{2}(z_{s}-z_{k})-\alpha\frac{dz_{s}}{dt}+O(\left\langle \sigma^{-}\sigma^{z}\sigma^{+}\right\rangle ).\label{eq: d2t with zkt}
\end{equation}
where $\left\langle \sigma^{-}\sigma^{z}\sigma^{+}\right\rangle $
denotes the last two terms in (\ref{eq: unit part}). The terms represented
by $\left\langle \sigma^{-}\sigma^{z}\sigma^{+}\right\rangle $ correspond
to off-diagonal elements in total density matrix. In particular, these
elements connect different spins (i.e., it is not just the coherence
of the system spin) as such, these elements are strongly suppressed
by the external dephasing, and can safely be neglected.

Note that in general $z_{k}=z_{k}(t)$ so this equation is not closed
and cannot be solved without prior knowledge of $z_{k}(t)$. There
are scenarios, though, in which this equation is closed, and can be
solved. One is the all to all coupling, and the other is weak coupling.
In the all to all coupling $\xi_{sk}=\xi$ so we can write

\begin{align}
\frac{d^{2}}{dt^{2}}z_{s} & =-2\xi^{2}Nz_{s}+2\xi^{2}\sum_{k}(z_{k})-\alpha\frac{dz_{s}}{dt}\nonumber \\
 & =-2\xi^{2}Nz_{s}+2\xi^{2}(z_{tot}-z_{s})-\alpha\frac{dz_{s}}{dt}\nonumber \\
 & =-2\xi^{2}(N+1)z_{s}-2\xi^{2}z_{tot}-\alpha\frac{dz_{s}}{dt}\nonumber \\
 & =-2\xi^{2}(N+1)(z_{s}-\frac{z_{tot}}{N+1})-\alpha\frac{dz_{s}}{dt}.
\end{align}
Finally, we get
\begin{equation}
\frac{d^{2}}{dt^{2}}z_{s}=-2N\xi^{2}\frac{N+1}{N}(z_{s}-z_{\infty})-\alpha\frac{dz_{s}}{dt}\label{eq: d2t z8 a2a}
\end{equation}
In the weak coupling limit $z_{1}=z_{2}=...=z_{N}\neq z_{s}$. As
before the average polarization is a conserved quantity so we can
write $z_{k}(t)=[(N+1)z_{\infty}-z_{s}(t)]/N$ and get

\begin{align}
\frac{d^{2}}{dt^{2}}z_{s} & =-2(\sum\xi_{sk}^{2})\frac{N+1}{N}(z_{s}-z_{\infty})-\alpha\frac{dz_{s}}{dt},\label{eq: d2t z8 NN}
\end{align}
which has the same form as (\ref{eq: d2t z8 a2a}) just with different
coupling strength. 

\subsection*{An extended dephasing time in small baths $T_{2}\le2\frac{N+1}{N}T_{1}$}

Consider a completely positive Markovian map (\ref{eq: Lindblad})
with $H=0$ for simplicity. $z_{0}$ is the asymptotic polarization
of the map. The polarization of the system $\left\langle \sigma^{z}\right\rangle $
can be changed by setting the Lindblad operators to $A_{1}=\sqrt{g_{+}}\sigma^{+},\:A_{2}=\sqrt{g_{-}}\sigma^{-}$.
We start by studying a map with time-independent asymptotic polarization
$\frac{dz_{0}}{dt}=0$ and get from (\ref{eq: Lindblad})
\begin{align}
\frac{d}{dt}z & =-2(g_{+}+g_{-})(z-z_{0}).\label{eq: z rate}\\
 & z_{0}=\frac{1}{2}\frac{g_{+}-g_{-}}{g_{+}+g_{-}},\label{eq: z0}
\end{align}
The coherence dynamics is obtained by looking on $\left\langle x\right\rangle =tr[\rho\sigma^{x}]$
that satisfies

\begin{equation}
\frac{d}{dt}x=-(g_{+}+g_{-})x.\label{eq: x rate}
\end{equation}
The solutions (\ref{eq: z rate}) and (\ref{eq: x rate}) are 
\begin{align}
\frac{z(t)-z_{0}}{z(0)-z_{0}} & =\exp[-2(g_{+}+g_{-})t]\triangleq e^{-\gamma_{z}t},\\
\frac{x(t)}{x(0)} & =\exp[-(g_{+}+g_{-})t]\triangleq e^{-\gamma_{x}t}.
\end{align}
Using these solution we compare the polarization decay rate $\gamma_{z}$
and the coherence decay rate $\gamma_{x}$ and get

\begin{equation}
\frac{\gamma_{z}}{\gamma_{x}}=2.
\end{equation}
Note that we have used the fact that $\frac{dz_{0}}{dt}=0$. Adding
elements like $A_{3}=\sigma_{z}$ will increase $\frac{d}{dt}x$ but
will not affect $\frac{d}{dt}z$ and therefore we get the standard
result for completely positive Markovian dynamics \cite{gorini76,chang1993T1T2}
\begin{equation}
\gamma_{z}\le2\gamma_{x}.
\end{equation}
In our case the map is \textit{time-dependent} . The rate $(g_{+}+g_{-})$
is fixed by the physical couplings, but $z_{0}$ (related to $g_{+}-g_{-}$)
changes in time since the bath is finite. Using the polarization conservation
$z_{0}(t)=\frac{(N+1)z_{\infty}-z_{s}}{N}$ we find 
\begin{equation}
\frac{d}{dt}z_{s}=-2(g_{+}+g_{-})\frac{N+1}{N}(z_{s}-z_{\infty}),
\end{equation}
and therefore 
\begin{equation}
\gamma_{z}\le2\frac{N+1}{N}\gamma_{x}.
\end{equation}
Thus, the polarization decay rate can be faster than the minimal value
of $2\gamma_{x}$ allowed for Markovian maps with a \textit{time-independent}
$z_{0}$. Note that this dressing effect does not happen for $x$
since the dephasing constantly eliminates any bath coherence that
may come from interacting with the system. 

\subsection*{Experimental realization in ion traps}

To synthesize significant pseudo-spin interaction Hamiltonians between
the ions in the trap, spin-dependent forces are used. These forces
can be realized, for example, by optical fields acting on optically
separated pseudo-spin levels, or by Raman transitions on microwave-separated
pseudo spin levels. Spin-dependent forces induce spin-dependent motion
of ions in the trap, leading to the acquisition of spin-dependent
phases, and thus to an effective spin-spin interaction. To mimic the
effect of spin-interaction Hamiltonians and not only their time-evolution
operator at specific times, quantum simulations are conducted using
spin-dependent forces that are tuned far off-resonance from one, or
more, of the crystal normal-modes of motion. Thus, excited motion
can be adiabatically eliminated and the interaction between the spins
becomes direct. Here, spin-spin interaction can be thought of as mediated
by the exchange of virtual crystal-phonons. A transverse field in
the $z$ direction can be introduced by detuning the pseudo-spin transition
from the Raman or optical interaction. The dephasing of bath spins
is straightforward to implement using individual-addressing of bath
ions with off-resonant laser beams that will shift them from resonance
in a quasi-random time sequence. 

The implementation of different $\xi_{i,j}$, depends on the normal
modes of motion that are used. This can be done by spectrally tuning
the lasers or microwave fields that induce spin-dependent forces.
ATA coupling can be achieved in ion-traps by tuning spin-dependent
forces close to the center-of-mass mode of an ion crystal in which
all ions oscillate in-phase, and with equal amplitude, along the trap
axis \cite{Schaetz2008}. Thus the ATA configuration discussed above
can be readily implemented. The use of spin-dependent forces that
act on radial normal modes in ions traps was shown to lead to relative
flexibility in determining $\xi_{i,j}$. In particular it was shown
that when radial modes are spectrally closely-spaced, the range of
spin-spin interactions can be scanned between $0\leq\delta\le3$ where
$\xi_{i,j}\propto1/|i-j|^{\delta}$ and $i$ and $j$ are the locations
of the ions in the chain, by tuning the spin-dependent force frequency
close to or far from the radial modes respectively \cite{Monroe2011}.
While the synthesis of arbitrary $\xi_{i,j}$ was shown to be possible
\cite{Monroe2012} it will be very difficult to experimentally implement.
The implementation of the NN configuration will therefore be challenging
using trapped-ion systems, although it could be fairly well approximated
using $\xi_{i,j}\propto1/|i-j|^{3}$ where the next-to-nearest neighbor
interaction is suppressed by a factor of eight. 

\subsection*{Experimental realization in Superconducting circuits}

For definiteness, we focus on superconducting qubits of the transmon
type \cite{Koch2007}. These qubits have good coherence properties,
can be individually addressed, made to interact, and read out with
high fidelity. Hence, they are currently being considered as building
blocks for quantum computation \cite{Devoret2013}. In particular,
they were recently used to study thermalization of an isolated quantum
system \cite{Neill2016}. The interaction between any pair $(i,j)$
of qubits can be realized using a common resonator as quantum bus
\cite{Blais2004,Blais2007} and takes the form $H_{I}=g_{i}g_{j}\left(\Delta_{i}^{-1}+\Delta_{j}^{-1}\right)\left(\sigma_{+,j}\sigma_{-,i}+\sigma_{+,i}\sigma_{-,j}\right)$,
where $g_{i}$ are the qubit-resonator couplings, $\Delta_{i}=\omega_{i}-\omega_{r}$
is the detuning between the qubit frequency $\omega_{i}$ and the
resonator frequency $\omega_{r}$, and the expression is valid in
the strong-detuning limit, $\Delta_{i}/g_{i}\gg1$. This interaction
realizes a two-qubit $\sqrt{i{\rm SWAP}}$ gate, mediated by virtual
interaction with the resonator. The interaction is effectively switched
off by ``parking'' the qubits in a largely detuned configuration,
and switched on by nonadiabatically tuning the qubits in resonance
with each other and closer in frequency to the resonator. (The qubit
frequencies can be adjusted on a fast time scale by changing the local
magnetic field at each qubit's site.) The NN scheme can be realized
by arranging the qubits in a one-dimensional chain and coupling each
neighboring pair by an individual bus resonator \cite{Salathe2015}.
The ATA scheme can be realized by coupling all qubits to a common
resonator, as recently demonstrated for an ensemble of ten qubits
in Ref. \cite{Song2017}. Based on these experiments, we estimate
that a tunable interaction strength $\xi/2\pi=5\text{MHz}$ can be
reached in both configurations. Single-qubit dephasing of arbitrary
strength can be engineered by injecting classical noise into the system,
causing fluctuations in the local magnetic field and hence in the
qubit frequency. Due to spurious coupling to uncontrolled degrees
of freedom the superconducting qubits have an intrinsic thermalization
time $T_{1}$. We require that the thermalization rate of the system
via the dephased bath be much larger than its intrinsic relaxation
rate, $\xi^{2}/\alpha\gg1/T_{1}$, so that the composite system can
be considered as isolated during the thermalization time. At the same
time, observing the full crossover between unitary dynamics and Zeno
freezing requires $\xi/\alpha\ll1$. Even assuming a conservative
$T_{1}\approx10~\mu{\rm s}$, ratios as low as $\xi/\alpha\approx1/20$
can be attained with intrinsic relaxation still playing a negligible
role.

\bibliographystyle{apsrev4-1}
\bibliography{/Users/raam_uzdin/Dropbox/RaamCite,ZenoBathExpRef}

\end{document}